# Minimizing the Switch and Link Conflicts in an Optical Multi-stage Interconnection Network

**Ved Prakash Bhardwaj[1], Nitin[2] and Vipin Tyagi[3]**

[1] **Department of Computer Science & Engineering and Information Technology, Jaypee University of Information Technology, Waknaghat, Solan-173234, Himachal Pradesh, India**

[2] **Department of Computer Science, College of Information Science and Technolopgy, University of Nebraska at Omaha, Omaha-68182-0116, Nebraska, United States of America**

[3] **Department of Computer Science & Engineering and Information Technology, Jaypee University of Engineering and Technology, Raghogarh, Guna-473226, Madhya Pradesh, India**

## Abstract

Multistage Interconnection Networks (MINs) are very popular in switching and communication applications. A MIN connects N inputs to N outputs and is referred as an N × N MIN; having size N. Optical Multistage Interconnection Network (OMIN) represents an important class of Interconnection networks. Crosstalk is the basic problem of OMIN. Switch Conflict and Link Conflict are the two main reason of crosstalk. In this paper, we are considering both problems. A number of techniques like Optical window, Improved Window, Heuristic, Genetic, and Zero have been proposed earlier in this research domain. In this paper, we have proposed two algorithms called Address Selection Algorithm and Route Selection Algorithm (RSA). RSA is based on Improved Window Method. We have applied the proposed algorithms on existing Omega network, having shuffle-exchange connection pattern. The main functionality of ASA and RSA is to minimize the number of switch and link conflicts in the network and to provide conflict free routes.

## 1. Introduction and Motivation

Parallel processing is an essential field for the present research environment, a v ital component of this environment is the Interconnection Network (IN). Multistage Interconnection Network is a low cost network, which interconnects N inputs with N outputs and has $\log_2 N$ switching stages [1-18]. MIN is used to develop more capable and cost effective systems. It provides low latency and can handle large amount of traffic efficiently [21, 22]. MIN can operate in the SIMD, multiple-SIMD, MIMD and partitionable SIMD/MIMD systems due to their full accessibility and reduced complexity, compared with crossbar switches [23]. It has a number of applications in the areas of computer and communications.

MIN is a very effective architecture for the multiprocessor systems as well as high speed communication systems [24]. It is also used in Optical data transmission. Optical MIN can work as an efficient approach in the Optical technology. So Optical MIN is the emerging field in MINs [21, 22].

Optical technology establishes new advancement to the field of communication network [25]. It is better than electronic communication in terms of bandwidth and latency. Optical Multi-stage Interconnection Network consists of N inputs, N outputs, and n stages (n=$\log_2 N$). Each stage has N/2 Switching Elements (SEs) comprising of two inputs and two outputs connected in a particular pattern [26]. It has many challenges like crosstalk, path dependent loss and load balancing. The problem of crosstalk may occur when two signals within a switch tries to interact with each other. The main reason of crosstalk is switch conflict and link conflict. Switch conflict can be tolerated. Now many devices are available to avoid switch conflicts like Noise re-correction device [27]. However, link conflict cannot be tolerated because two messages cannot traverse along the same link at the same time [28].

In the present paper, our interest is on the Time Domain Approach (TDA) for solving switch and link conflict in Optical Omega Network (OON) [29]. In this approach, the two communication signals will be passed at different times if they are using same switching element. In this paper, we have proposed two algorithms called Address Selection Algorithm and Route Selection Algorithm. Both are minimizing the switch and link conflicts in the network and provides crosstalk free routes in the network [30].

## 2. Crosstalk

The problem of crosstalk may occur when two signals within a switch tries to interact with each other. There are various reasons of crosstalk which will be explained in this section.

### 2.1 Problem of Switch Conflict in Optical Omega Network

The Omega Network (ON) is an example of a banyan multistage interconnection network [31]. This network has a shuffle exchange connection pattern. In this pattern the address is shifted one bit to the left circularly in each connection. This network connects N input to N output nodes using n stages, where $n = \log_2 N$ with each stage containing $2^{n-1}$ SEs. In the Optical Omega Network , each communication signal must go through a number of switching stages. Each data packet is having a specific data path according to its destination. In switch conflict problem [32] the two source addresses are trying to interact with each other within a same switching element. Figure (1) shows a 3-stage, 8 x 8 Optical Omega Network. In this network all the source addresses are getting their destination in a single time slot [33], so it creates the problem of switch conflict in the network. It is shown by arrows in figure (1). The arrows in figure (1) indicate the conflicted routes. There are nine conflicts in the network. All conflicted routes having a common reason viz. the communication signals are trying to interact with each other within a switching element.

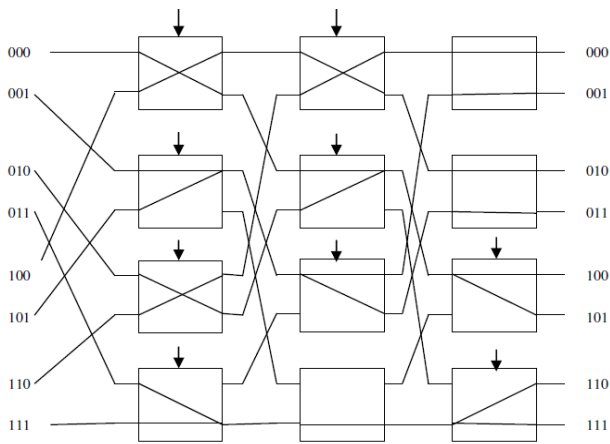

Fig.1 An 8×8 Optical Omega Network With Switch Conflicts.

### 2.2 Problem of Link Conflict in Optical Omega Network

Each data packet is having a specific data path according to its destination. In link conflict problem [34] the two source addresses are trying to traverse on the same path in a same time slot. Practically this activity is not possible. Figure (2) shows a 3-stage, 8 x 8 Optical Omega Network. In this network all the source addresses are getting their destination in a single time slot [33], so it creates the problem of link conflict in the network. It is shown by dotted lines in figure (2). The arrows in figure (2) indicate the conflicted routes. There are four conflicts in the network. All conflicted routes having a common reason viz. the communication signals are trying to move on the same path at the same time.

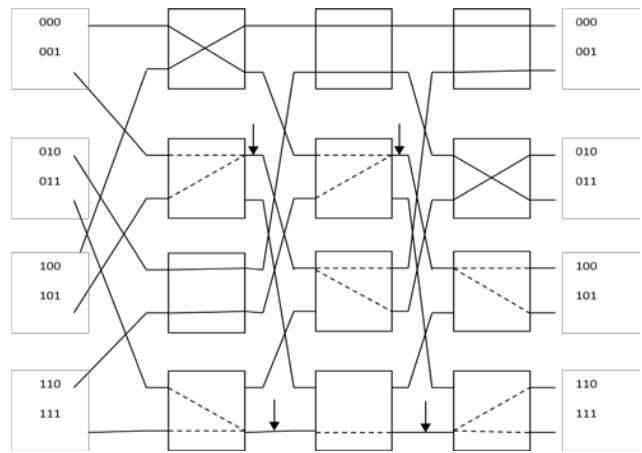

Figure2: An 8×8 Optical Omega Network With Link Conflicts.

## 3. Preliminaries and Background

There are various approaches available to reduce the problem of crosstalk like space domain, time domain and wavelength domain approach. In the present paper, our consideration is time domain approach [25]. Crosstalk is considered as a conflict in this approach. It is a good approach because it can make a balance between the electronic processor and Optical MINs [26, 27]. It is not possible to send all the source addresses at the same time to their corresponding destination because it will create the switch and link conflict problem. Therefore, to route the data packets, Permutation and Semi-permutation is applied on the message groups. So that a conflict free route can be obtained for each group [28]. The source and destination address is combined to build combination matrix. On the basis of combination matrix message partitioning is performed so that some specific message should get their destination in the first pass and network remains crosstalk free. There are various techniques for message partitioning like Window Method [29], Improved Window Method [30] and Heuristic Routing Algorithms [30]. In this paper, the focus is to provide best message partitioning scheme so that a switch and link conflict free network can be obtained. Before describing our algorithms just have a look on the Window Method and Improved Window Method and Heuristic Routing Algorithm.

## 3.1 Window Method

This method [28] basically separates the messages, which have the same bit pattern so that crosstalk can be removed. If we consider the network size N x N, it shows that there are N source and N destination address. To get a combination matrix, it is required to combine the corresponding source and destination address. This matrix shows that the optical window size is M-1, where M=log2N and N is the size of the network. The first and last columns of the combination matrix are not considered in this method and all the processing is performed on the remaining columns. If the two messages having the same bit pattern then they will be routed in the different passes.

## 3.2 Improved Window Method

The first window is not considered in this method and all the processing is performed on the remaining windows [28, 29]. It is faster than the window method. The conflict matrix is obtained on the basis of remaining windows [30, 31]. This is the pseudo code of the IWM algorithm.

```
Improved Window Method()
Begin
 Initialize conflict matrix with zero;
 for(n=0 to N/2-1) do
    Set conflict Matrix [n] [n+N/2] with 1
 End for;
    Window()
 End for;
End;
```

## 3.3 Heuristic Routing Algorithm

There are four approaches of this algorithm to schedule the messages in different passes in order to avoid the path conflicts in the network [27]. In the first approach, the message is selected on the basis of their increasing order. In second approach, the message is selected on the basis of their decreasing order. In third approach, the main consideration is on the degree of the conflict graph. The message with the lowest degree will be selected first; similarly the other message can be selected. Finally in the fourth approach, the message with the highest degree will be selected first; similarly, the other message can be selected.

## 4. Proposed Algorithms

The Address Selection Algorithm is minimizing the switch conflict of the network and Route Selection Algorithm is minimizing the link conflict of the network.

## 4.1 Address Selection Algorithm

This algorithm is based on the above work. The aim of this algorithm is to select such particular source address in first pass, which do not create conflict in the network, and the remaining source address can be transmitted in second pass. This algorithm is applicable on 8 x 8 Optical Multistage Interconnection Network and its above series. In our approach first, we get the source and destination address sequentially. Second, we find the combination matrix of the source and corresponding destination address. Now transformation is applied on the combination matrix. The transposed matrix will have a particular set of rows. Now select the middle four rows and eliminate the remaining rows. In this way, two pair of rows can be obtained. In the next step, addition operation will be performed between corresponding bits in each pair. Therefore, two different sets will be obtained for the next step. We subtract the obtained result of first set with its corresponding result of second set.

**Address_Selection Algorithm (ASA)**

1. Get the source and destination address sequentially.
2. Make combination matrix of the source and corresponding destination address.
3. Transform the matrix. A complete set of rows are thus formed $r_0$, $r_1....r_n$, where n=total number of bits in source and destination address.
4. Select the middle four rows.
5. Pair the obtained rows in set of two.
6. Add the corresponding bits of each pair.
7. Subtract the result of first set from second set.
8. If(result <= 0)
      Then take corresponding address and transmit them in current pass and go to step 9.
   Else
      Store the address in remaining_address.
9. If(Conflict)
      Then transmit the address with higher magnitude of the conflicting address pair and add the lower magnitude address to the remaining_address.
10. Transmit the remaining_address.
11. End.

If the result is a positive number then store it in a variable called the remaining_address. If conflict occurs in the current pass then we transmit only those addresses, which have higher magnitude and the address, which has lower magnitude, store it in the remaining_address. Now in the second pass, transmit the all remaining addresses, which are store in remaining_address. In this way, a conflict free network can be obtained.

## 4.2 Route Selection Algorithm

The aim of this algorithm is to select such particular source address in first pass, which do not create link conflict in the network, and the remaining source address can be transmitted in second pass. This algorithm is applicable on $8 \times 8$ Optical Multistage Interconnection Network and its above series. In our approach first, we get the source and destination address sequentially. Second, we find the combination matrix of the source and corresponding destination address. Now middle four columns will be selected from the combination matrix. To get the conflict matrix [31], Improved Window Method [29, 30] is applied on the selected columns. Now row wise addition is applied in the matrix so that sum of each row can be obtained. If the sum is zero then it will be stored in selected_list variable otherwise store the addresses in conflicted_list. Now select the source addresses from conflicted_list, which are having magnitude equal to top two magnitude of the conflicted_list and store these addresses in the selected_list. Now we will transmit the selected_list addresses in the network.

**Route_Selection Algorithm (RSA)**

```
1. Get  the source and destination
   address sequentially.
2. Make combination matrix of the source
   and    corresponding destination
   address.
3. Select the middle four columns.
4. Apply  Improved Window Method on the
   selected columns to get the conflict
   matrix and apply row wise addition in
   the matrix for each row and get the
   sum.
5. If(sum = 0)
      Then take corresponding address
      and store it in selected_list.
   Else
      Store the address in
      conflicted_list.
6. Select  the source addresses, which
   are having magnitude equal to top two
   magnitude of the conflicted_list and
   store  these addres ses  in the
   selected_list.
7. Transmit the selected_list.
8. If(link_Conflict)
      Then transmit the address with
      higher magnitude of the
      conflicting address
      pair and add the lower magnitude
```

address to the conflicted_list.
```
9. Transmit the conflicted_list.
10. End.
```

If any link conflict occurs in the current pass then we transmit only those addresses, which have higher magnitude and the address, which has lower magnitude, store it in the conflicted_list. Now in the second pass, transmit the all remaining addresses, which are store in conflicted_list. In this way, a conflict free network can be obtained. This algorithm is minimizing the number of link conflict in the network.

## 5. Results

### 5.1 Address Selection Algorithm

Example1: Let the source and destination address as follows and these are going to be inputs for the Algorithmic steps.

| Source | Destination |
|--------|-------------|
| 000 | 100 |
| 001 | 011 |
| 010 | 101 |
| 011 | 110 |
| 100 | 010 |
| 101 | 001 |
| 110 | 000 |
| 111 | 111 |

Algorithmic Step1: Get the source and destination address sequentially.

| Source | Destination |
|--------|-------------|
| 000 | 100 |
| 001 | 011 |
| 010 | 101 |
| 011 | 110 |
| 100 | 010 |
| 101 | 001 |
| 110 | 000 |
| 111 | 111 |

Algorithmic Step2: Make combination matrix of the source and destination address. Like source address is 000 and destination address is 100. Therefore, it will become 000100. The other combination of source and destination address can be made on the same pattern. It is clear from the combination matrix.

$$\begin{bmatrix} 0 & 0 & 0 & 1 & 0 & 0 \\ 0 & 0 & 1 & 0 & 1 & 1 \\ 0 & 1 & 0 & 1 & 0 & 1 \\ 0 & 1 & 1 & 1 & 1 & 0 \\ 1 & 0 & 0 & 0 & 1 & 0 \\ 1 & 0 & 1 & 0 & 0 & 1 \\ 1 & 1 & 0 & 0 & 0 & 0 \\ 1 & 1 & 1 & 1 & 1 & 1 \end{bmatrix}$$

**Algorithmic Step3:** Transpose the matrix. Row 1 of the matrix represents $r_0$, similarly the other row of the matrix represents $r_1$ till $r_5$.

$$\begin{bmatrix} 0 & 0 & 0 & 0 & 1 & 1 & 1 & 1 \\ 0 & 0 & 1 & 1 & 0 & 0 & 1 & 1 \\ 0 & 1 & 0 & 1 & 0 & 1 & 0 & 1 \\ 1 & 0 & 1 & 1 & 0 & 0 & 0 & 1 \\ 0 & 1 & 0 & 1 & 1 & 0 & 0 & 1 \\ 0 & 1 & 1 & 0 & 0 & 1 & 0 & 1 \end{bmatrix}$$

**Algorithmic Step4:** select the middle four rows, eliminate the first and last row, and get the remaining rows. Row 1 of the matrix represents $r_1$, similarly the other row of the matrix represents $r_2$ till $r_4$.

$$\begin{bmatrix} 0 & 0 & 1 & 1 & 0 & 0 & 1 & 1 \\ 0 & 1 & 0 & 1 & 0 & 1 & 0 & 1 \\ 1 & 0 & 1 & 1 & 0 & 0 & 0 & 1 \\ 0 & 1 & 0 & 1 & 1 & 0 & 0 & 1 \end{bmatrix}$$

**Algorithmic Step5:** Pair the obtained rows in set of two. The first pair will have rows $r_1$ and $r_2$. The second pair will have rows $r_3$ and $r_4$.

| $r_1$ | $r_2$ | $r_3$ | $r_4$ |
|---|---|---|---|
| 0 | 0 | 1 | 0 |
| 0 | 1 | 0 | 1 |
| 1 | 0 | 1 | 0 |
| 1 | 1 | 1 | 1 |
| 0 | 0 | 0 | 1 |
| 0 | 1 | 0 | 0 |
| 1 | 0 | 0 | 0 |
| 1 | 1 | 1 | 1 |

**Algorithmic Step6:** Add the corresponding bits of each pair.

| $r_1+r_2$ | $r_3+r_4$ |
|---|---|
| 0 | 1 |
| 1 | 1 |
| 1 | 1 |
| 2 | 2 |
| 0 | 1 |
| 1 | 0 |
| 1 | 0 |
| 2 | 2 |

**Algorithmic Step7:** Subtract the result of the first set from the second set.

$(r_1+r_2)-(r_3+r_4)$
→ -1
→ 0
→ 0
→ 0
→ -1
　1
　1
→ 0

**Algorithmic Step8:** Now the corresponding addresses, having their result either zero or negative number will be transmitted. These addresses are 000, 001, 010, 011, 100 and111. The transmission of these selected addresses is clear from figure (3).

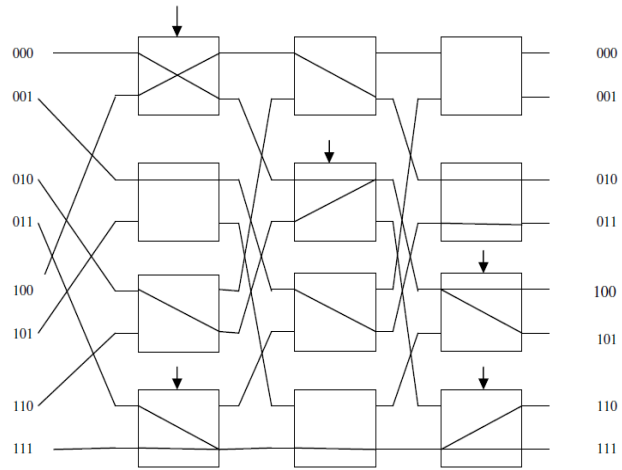

Fig.3 Optical Omega Network with Reduced Switch Conflict.

Now the addresses, which are having results in positive numbers, will be stored in remaining_address variable. These addresses are 101 and 110.

**Algorithmic Step9:** Now finding the conflicted addresses. From figure (3), the arrows show the conflict in the network. Therefore, select the conflicting addresses and find which is having greater magnitude. Therefore, in the above network 000 a nd 100 is having conflict. The addresses 011 a nd 111 h aving conflicts. Similarly, the other conflicted address can be obtained as shown in figure (3). Finally, store the address 000 a nd 011 in the remaining_address variable. In this way, the addresses 001, 010, 100 and 111 will be transmitted as shown in figure (4). It is clear from the figure that there is no conflict in the network.

**Algorithmic Step10:** remaining_address variable will store four source addresses for next pass i.e. 101, 110, 000 and

011. Similarly, the omega network for addresses 101, 110, 000 and 111 can be obtained in second pass. These addresses will get their destination without any conflict. So that the network remains conflict free.

Algorithmic Step11: Finally, goal of the algorithm is achieved.

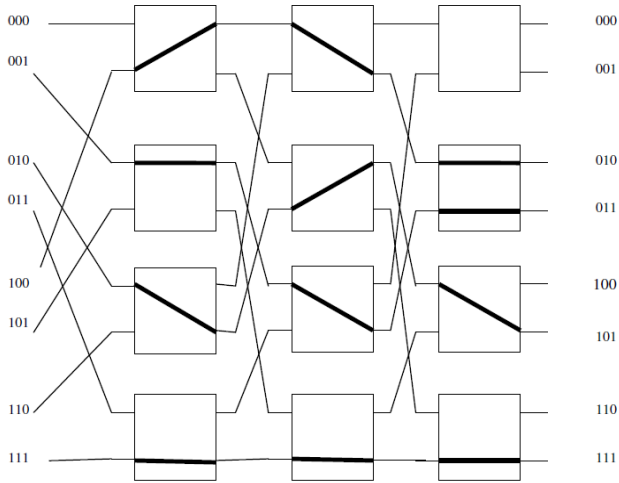

Fig.4 Conflict Free Optical Omega Network.

## 5.2 Route Selection Algorithm

Example1: Let the source and destination address as follows and these are going to be inputs for the Algorithmic steps.

| Source | Destination |
|--------|-------------|
| 000 | 101 |
| 001 | 001 |
| 010 | 011 |
| 011 | 110 |
| 100 | 000 |
| 101 | 010 |
| 110 | 100 |
| 111 | 111 |

Algorithmic Step1: Get the source and destination address sequentially.

| Source | Destination |
|--------|-------------|
| 000 | 101 |
| 001 | 001 |
| 010 | 011 |
| 011 | 110 |
| 100 | 000 |
| 101 | 010 |
| 110 | 100 |
| 111 | 111 |

Algorithmic Step2: Make combination matrix of the source and destination address. Like source address is 000 and destination address is 101. Therefore, it will become 000101. The other combination of source and destination address can be made on the same pattern. It is clear from the combination matrix.

$$\begin{bmatrix} 0 & 0 & 0 & 1 & 0 & 1 \\ 0 & 0 & 1 & 0 & 0 & 1 \\ 0 & 1 & 0 & 0 & 1 & 1 \\ 0 & 1 & 1 & 1 & 1 & 0 \\ 1 & 0 & 0 & 0 & 0 & 0 \\ 1 & 0 & 1 & 0 & 1 & 0 \\ 1 & 1 & 0 & 1 & 0 & 0 \\ 1 & 1 & 1 & 1 & 1 & 1 \end{bmatrix}$$

Algorithmic Step3: Select the middle four columns.

$$\begin{bmatrix} 0 & 0 & 1 & 0 \\ 0 & 1 & 0 & 0 \\ 1 & 0 & 0 & 1 \\ 1 & 1 & 1 & 1 \\ 0 & 0 & 0 & 0 \\ 0 & 1 & 0 & 1 \\ 1 & 0 & 1 & 0 \\ 1 & 1 & 1 & 1 \end{bmatrix}$$

Algorithmic Step4: Now we will apply the Improved Window Method [30] on the selected columns to get the conflict matrix [29, 30]. Window one will contain the column two and column three and Window two will contain the column three and column four.

| Window1 | Window2 |
|---------|---------|
| 0 1 | 1 0 |
| 1 0 | 0 0 |
| 0 0 | 0 1 |
| 1 1 | 1 1 |
| 0 0 | 0 0 |
| 1 0 | 0 1 |
| 0 1 | 1 0 |
| 1 1 | 1 1 |

Table 1: Conflict Matrix Table

| msg | 000 | 001 | 010 | 011 | 100 | 101 | 110 | 111 | sum |
|-----|-----|-----|-----|-----|-----|-----|-----|-----|-----|
| 000 | 0 | 0 | 0 | 0 | 0 | 0 | 1 | 0 | 1 |
| 001 | 0 | 0 | 0 | 0 | 1 | 1 | 0 | 0 | 2 |
| 010 | 0 | 0 | 0 | 0 | 1 | 1 | 0 | 0 | 2 |
| 011 | 0 | 0 | 0 | 0 | 0 | 0 | 0 | 1 | 1 |
| 100 | 0 | 0 | 0 | 0 | 0 | 0 | 0 | 0 | 0 |
| 101 | 0 | 0 | 0 | 0 | 0 | 0 | 0 | 0 | 0 |
| 110 | 0 | 0 | 0 | 0 | 0 | 0 | 0 | 0 | 0 |
| 111 | 0 | 0 | 0 | 0 | 0 | 0 | 0 | 0 | 0 |

Algorithmic Step5: From the conflict matrix [31, 32] table it is clear that the source addresses 100,101,110 and 111 having summed zero. Therefore, we store these addresses in selected_list variable. The remaining source addresses will be stored in the conflicted_list variable. These are 000,001,010, and 011.

Algorithmic Step6: So in this step we will select 010 and 011 from the conflicted_list because these addresses are high in magnitude comparison to the other addresses. Store the selected addresses in the selected_list variable. Now the selected_list will have 100,101,110,111,010 and 011 for the transmission.

Algorithmic Step7: Now we will transmit the selected_list. It is shown in figure (5).

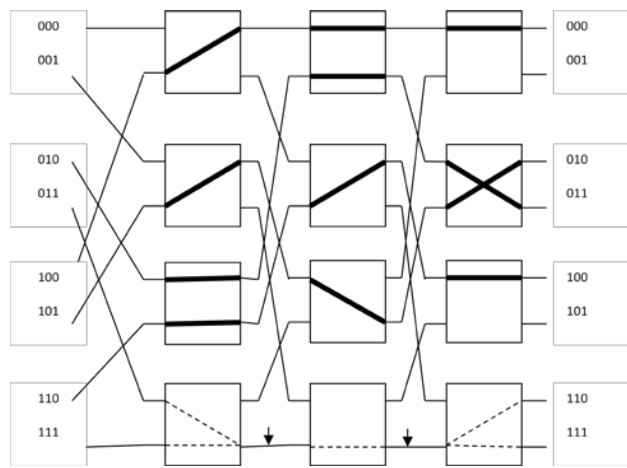

Fig.5 An 8×8 Optical Omega Network With Reduced Link Conflicts.

Algorithmic Step8: From figure (5) it is clear that there is a link conflict between the source address 011 and 111 and 111 is higher in magnitude comparison to 011. So transmit 111 only in the current pass. 011 will be stored in conflict_list.

Algorithmic Step9: So, the conflict_list will have the addresses 000,001 and 011 for the second pass. Similarly, we can obtain the second pass by transmitting the remaining addresses. It is clear from the figure (6).

Algorithmic Step10: Our goal is to minimize the link conflict in the omega network. In figure (2) the number of link conflict is four and in figure (5) the number of link conflict is two. It is shown by dotted lines. This algorithm can reduce more than 50% link conflict of the network. Finally, goal of the algorithm is achieved.

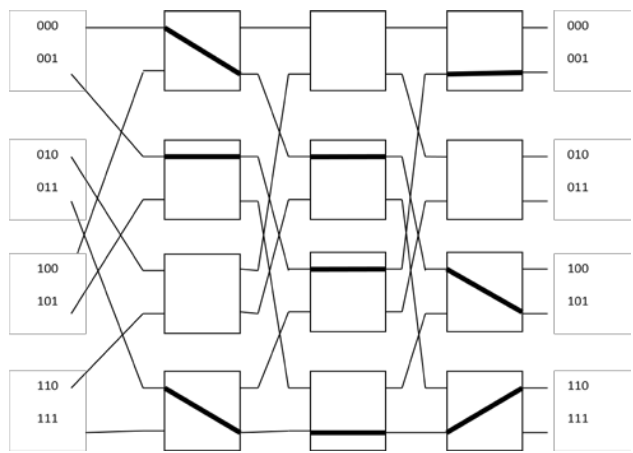

Figure 6: An 8×8 Optical Omega Network With Link Conflict Free Routes.

## 6. Conclusion

Optical Multistage Interconnection Network is the advance field in communication technology. Crosstalk is the biggest challenge in the Optical Multistage Interconnection Network. It is created because of Switch and Link Conflict. Switch Conflict can be tolerated because of advances in Optical Technology but in case of link conflict the two source addresses try to move on the data path at the same time. Practically, it is not possible. In this paper, we have presented two algorithms called Address Selection Algorithm and Route Selection Algorithm. Both algorithms select some specific source addresses for first pass and these addresses will be transmitted through the optical omega network. In the second pass, it transmits the remaining source address to their destination. In this, way both algorithms are minimizing the switch and link conflicts of the network. Therefore, both proposed algorithms can work as a solution to avoid Switch and Link conflict in OMIN. These approaches can be applied to other Time Domain Algorithms.

## Appendix

Here we are presenting the code we have design for implementing our results.

```
//Initialization
```

```
int c[][]=new int[6][8];
nt temp1[]=new int[8];
int temp2[]=new int[8];
int result[]=new int [8];
InputStreamReader fm=new
InputStreamReader(System.in);
BufferedReader br=new BufferedReader(fm);

for(int k=0;k<6;k++)
  for(int l=0;l<8;l++)
    c[k][l]=Integer.parseInt(br.readLine());

//performing computation
for(int k=0;k<6;k++)
  for(int l=0;l<8;l++)
    temp1[l] = c[1][l]+c[2][l];
    temp2[l] = c[3][l] +c[4][l];

//computing result
for(int l=0;l<8;l++)
  result[l]=temp1[l]-temp2[l];

//displaying the result
for(int l=0;l<8;l++)
  System.out.println("C1+C2 : " +temp1[l]);
for(int l=0;l<8;l++)
  System.out.println("C3+C4 : " +temp2[l]);
for(int l=0;l<8;l++)
  System.out.println("Result : " +result[l]);
```

**Ved Prakash Bhardwaj** Completed the Bachelor of Engineering in Computer Engineering in the year 2008 from Rajasthan University, Jaipur, Master of Technology in Computer Science in the year 2010 from Jaypee University of Information Technology, Solan. Presently doing Ph.D. in Computer Science from Jaypee University of Information Technology, Solan. The research interest include Interconnection Networks, Multistage Interconnection Network.

**Dr. Nitin** is currently working as a Distinguished Adjunct Professor in the Department of Computer Science, College of Information Science and Technology, The Peter Kiewit Institute, University of Nebraska at Omaha, Omaha, NE, USA. His permanent affiliation is with the Department of Computer Science & Engineering and Information Technology, Jaypee University of Information Technology (JUIT), Waknaghat, Solan-173234, Himachal Pradesh, India. He was born on October 06, 1978. He received his Bachelor's degree in Computer Science & Engineering [Hons.] in 2001 and Master's Degree in Software Engineering from Thapar Institute of Engineering and Technology, Patiala, India in year 2003. In 2009, he received his Doctor of Philosophy degree in Computer Science & Engineering from JUIT. He is a IBM certified engineer. He is a Life Member of IAENG, Senior Member IACSIT and Member of SIAM, IEEE and ACIS and have 71 research papers in peer reviewed International Journals & Transactions, Book Chapters and Conferences. His research interest includes Interconnection Networks & Architecture, Fault-tolerance & Reliability, Networks-on-Chip, Systems-on-Chip, and Networks-in-Packages, Application of Stable Matching Problems, Stochastic Communication and Graph Embedding, Underwater & Sensor Networks. Currently he is working on Parallel Simulation tools, BigNetSim using Charm++, NS-2 using TCL. He is the Co-founder of High-end Parallel Computing and Advanced Computer Architecture Lab at JUIT. Recently he has been appointed as a Co-editor, International Journal of Computer Theory and Engineering (IJCTE), Singapore, International Journal of Advancements in Computing Technology (IJACT), Korea. He is also referee for the [Mathematical and Computer Modelling, Journal of Parallel and Distributed Computing, Computer Communications, Computers and electrical Engineering]@Elsevier Sciences, WSEAS Transactions, The Journal of Supercomputing, Springer and International Journal of System Science, Taylor & Francis.

**Dr. Vipin Tyagi is** Associate Prof. in Computer Science and Engineering at Jaypee University of Engg and Technology, Raghogarh , Guna (MP) India. He holds Ph.D. degree, M.Tech. in Computer Science and Engineering and MSc in Mathematics. He has about 20 years of teaching and research experience. He is an active member of Indian Science Congress Association and President of Engineering Sciences Section of the Association for the term 2010-11. He is a senior life member of Computer Society of India. He is member of Academic-Research and Consultancy committee of Computer Society of India. He is elected as Executive Committee member of Bhopal Chapter of Computer Society of India and M.P. and C G chapter of IETE. He is state student coordinator of MP state of Computer Society of India. He is a Fellow of Institution of Electronics and Telecommunication Engineers, life member of CSI, Indian Remote Sensing Society, CSTA, ISCA and IEEE, International Association of Engineers.